# Direct vs. indirect optical recombination in Ge films grown on Si substrates

G. Grzybowski[1], R. Roucka[1], J. Mathews[2], R.T Beeler[1], J. Kouvetakis[1] and J. Menéndez[2]

[1]Department of Chemistry and Biochemistry, Arizona State University, Tempe, AZ 85287-1604

[2]Department of Physics, Arizona State University, Tempe, AZ 85287-1504

**Abstract:** The optical emission spectra from Ge films on Si are markedly different from their bulk Ge counterparts. Whereas bulk Ge emission is dominated by the material's indirect gap, the photoluminescence signal from Ge films is mainly associated with its direct band gap. Using a new class of Ge-on-Si films grown by a recently introduced CVD approach, we study the direct and indirect photoluminescence from intrinsic and doped samples and we conclude that the origin of the discrepancy is the lack of self-absorption in thin Ge films combined with a deviation from quasi-equilibrium conditions in the conduction band. The latter is confirmed by a simple model suggesting that the deviation from quasi-equilibrium is caused by the much shorter recombination lifetime in the films relative to bulk Ge.

Germanium has unique optical properties related to the small separation of 0.14 eV between the absolute minimum of the conduction band at the *L*-point of the Brillouin zone and the Γ-point local minimum. Possible perturbation schemes leading to direct or quasi-direct band gap conditions have been known for a long time. These include *n*-type doping,[1] alloying with Sn,[2] and the application of tensile strain.[3] Over the past few years these approaches have been contemplated as a possible way to integrate direct gap semiconductors with Si technology, fueling an intense research effort that culminated in the recent announcement of an optically pumped Ge-on-Si laser with emission close to 1550 nm.[4]

A remarkable feature of all published room temperature luminescence studies of Ge-on-Si films is the dominance of direct gap emission from the Γ-point local minimum in the conduction band to the absolute maximum of the valence band at the same $k = 0$ wave vector.[5-10] This is in sharp contrast with similar experiments on bulk Ge samples,[11] for which indirect gap emission from the *L*-minimum is much stronger. In this communication, we present a systematic study focused on explaining this discrepancy. We believe that the understanding of this puzzle may contribute to the advancement of this field towards the final goal of developing electrically-injected lasers integrated on Si substrates.

We studied Ge-on-Si samples grown by a novel CVD approach recently introduced by Beeler and co-workers.[12] This approach is an extrapolation—to ultra-low Sn concentrations—of the growth procedure introduced by Bauer and co-workers to synthesize $Ge_{1-y}Sn_y$ alloys.[13] Briefly, the growth is conducted on high resistivity Si(100) wafers via reactions of digermane ($Ge_2H_6$) and deuterated stannane ($SnD_4$) diluted in $H_2$. Films with thicknesses up to several microns are commonly produced at high growth rates up to 30 nm/min via reactions of $Ge_2H_6$ with appropriate amounts of $SnD_4$ at $T$=380-400 $^oC$. The incorporation of dopant levels of substitutional Sn into the Ge-on-Si films at nominal levels of 0.03-0.15% is sufficient to completely suppress the traditional layer-plus-island growth mode (Stranski-Krastanov). The resultant layers are found to exhibit flat surfaces, fully relaxed strain states, and crystallinity/morphology comparable to those observed in the intrinsic materials, as evidenced by a range of analytical methods including Rutherford Backscattering, Atomic Force Microscopy, high resolution X-ray Diffraction, Secondary Ion Mass Spectrometry and Cross-sectional Transmission Electron Microscopy. The small Sn-concentrations do not alter the optical properties in any measurable way, thus the material is referred to as "quasi-Ge". The films can be systematically co-doped with P atoms at controlled levels of up to $2 \times 10^{19}$ /$cm^3$ in situ using the single-source $P(GeH_3)_3$. This process produces tunable and highly controlled atomic profiles of the donor atoms by judiciously adjusting the $P(GeH_3)_3/Ge_2H_6$ ratio in the reaction mixture.

Photoluminescence (PL) was measured using a 980 nm laser focused to a ~100 μm spot on samples with thicknesses on the order of 1μm. The average incident power was 200 mW. The emitted light was analyzed with an $f$ = 320 mm spectrometer equipped with a 600 lines/mm grating blazed at 2 μm, and detected with a single-channel, $LN_2$-cooled extended InGaAs receiver (1.3-2.3 μm range). Figure 1 shows the PL signal from a "quasi-Ge" film on Si. The as-grown sample shows no measurable PL, but after rapid thermal annealing (RTA) at 680°C (which improves crystallinity, as seen from the narrowing of the x-ray diffraction peaks), followed by a heat treatment under hydrogen, as described in Ref. 8, the signal is maximized. The RTA treatment causes a tensile strain due to the thermal expansion mismatch with the Si substrate. In the case of the sample in Fig. 1, this strain is $\varepsilon = 0.24\%$. The main emission feature is assigned to the direct gap $E_0$. The shoulder at lower energies corresponds to indirect gap emission. The PL spectrum is virtually identical to that from a pure Ge film on Si (not shown) grown by the method introduced by Wistey et al.(Ref. 14) This confirms that the extremely small amount of Sn in the "quasi-Ge" material does not change the optical properties in any measurable way. Accordingly, we will simply refer to our films as Ge-on-Si, ignoring the small residual Sn concentration.

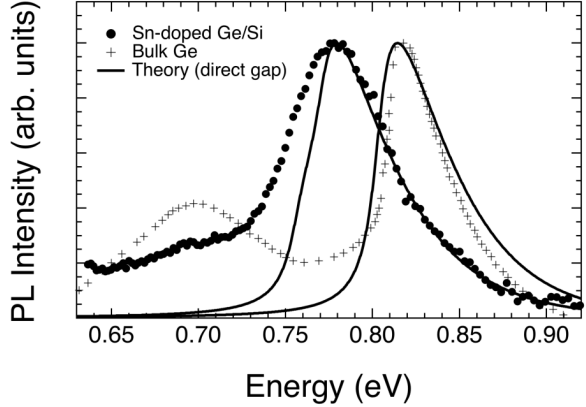

**Figure 1** Room temperature photoluminescence spectrum of "quasi-Ge" compared with the corresponding spectrum from bulk Ge after self-abosorption corrections (Ref. 17). The maximum is assigned to direct gap emission. The weaker feature at lower energy corresponds to indirect gap emission. The spectra are normalized to the same peak intensity. The solid lines are theoretical fits to the direct gap emission, as discussed in the text.

It seems natural to compare the film PL with bulk Ge emission. However, this comparison must be done with care because the PL spectrum from bulk Ge is severely distorted by self-absorption effects. The diffusion length of electron-hole pairs in Ge exceeds 0.4 mm,[15, 16] so that reabsorption cannot be neglected, and since the absorption coefficient is higher for 0.8 eV than for 0.7 eV photons, the observed direct gap PL is preferentially suppressed. By performing experiments on thin specimens in a transmission geometry with lamp illumination, for which reabsorption can be easily computed, Haynes[17] generated a "corrected" emission spectrum for bulk Ge, which is shown as crosses in Fig. 1. This spectrum shows clear peaks corresponding to direct and indirect transitions, with the direct gap emission being the strongest one. Since our Ge-on-Si films have thicknesses on the order of 1 μm, self-absorption corrections are much less important and the corresponding spectra should be compared with the "corrected" one from bulk Ge. Thus the dominance of the direct gap emission in our Ge films is to be expected. However,

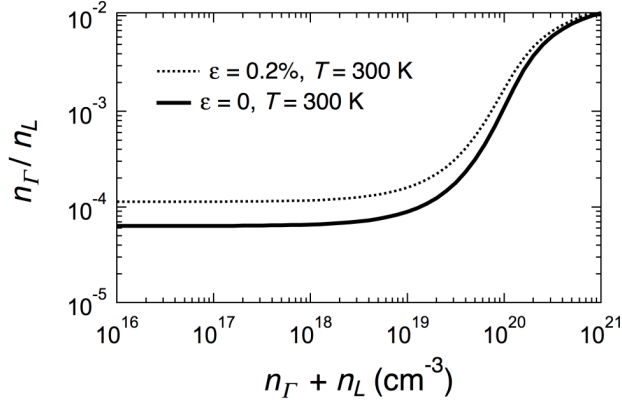

**Figure 2** Population ratio for the Γ and L valleys in the conduction band of Ge as a function of the total electron density in the conduction band. The calculations are shown for relaxed Ge ($\varepsilon = 0$) as well as for Ge with a biaxial tensile strain in the (001) plane ($\varepsilon = 0.2\%$)

self-absorption alone is insufficient to explain the experimental findings because the intensity ratio (peak areas) between the direct and indirect emission is $I_{dir}/I_{ind} = 8.5$ for the Ge film, but only $I_{dir}/I_{ind} = 1.3$ for bulk Ge. This relative suppression of indirect gap emission is observed quite systematically in all of our Ge-like films on Si (including $Ge_{1-y}Sn_y$) and has been noted earlier,[5-10] although its quantification is difficult when the PL is measured with standard Ge or InGaAs detectors, which cut off the indirect gap emission.

An additional discrepancy between the Ge-film emission and the corrected bulk Ge emission is the relative spectral redshift of the Ge film spectrum, as seen in Fig. 1. This shift can be understood in terms of strain and temperature. For this we model the direct gap PL following the method described in Ref. 9. The results are shown as solid lines in Fig. 1. For the bulk Ge case, we find a good agreement in peak position and overall lineshape assuming that the sample temperature is $T = 290$ K and the photoexcited electron density is less than $1 \times 10^{18}$ cm$^{-3}$. The residual disagreement between theory and experiment may be due to systematic errors in the self-absorption correction, caused mainly by the fact that the absorption coefficient of Ge in this spectral range is very strongly dependent on temperature.[18] For the Ge-on-Si sample, the emission is calculated using the measured tensile strain of 0.24% and a sample temperature of 320 K. The tensile strain explains 80% of the red shift with respect to bulk Ge, whereas the higher sample temperature, which is expected for laser excitation, accounts the remaining 20%.

The calculated direct gap emission requires as an input the carrier density $n_\Gamma$ at the Γ-valley. We use as an adjustable parameter the total photoexcited electron concentration $n_\Gamma + n_L$, where $n_L$ is the carrier density at the L-valley, and obtain $n_\Gamma$ by assuming quasi-equilibrium conditions between the two valleys. The model does not include indirect gap emission, so that we do not have a predicted value for $I_{dir}/I_{ind}$, but we can predict *changes* to this ratio, which is expected to be proportional to $n_\Gamma/n_L$.[19] The calculated $n_\Gamma/n_L$ is shown in Fig. 2 as a function of the total electron population. We see that $n_\Gamma/n_L$ is fairly constant in the $10^{16}$-$10^{19}$ cm$^{-3}$ concentration range, suggesting that the different $I_{dir}/I_{ind}$ ratios cannot be explained in terms of different photoexcitation levels. The ratio does increase with temperature and with tensile strain (which decreases the energy separation between the two valleys). For the temperature and strain corresponding to our Ge-on-Si sample, the $I_{dir}/I_{ind}$ ratio should have increased by a factor of 3 relative to bulk Ge, whereas the observed increase factor is larger than 6.

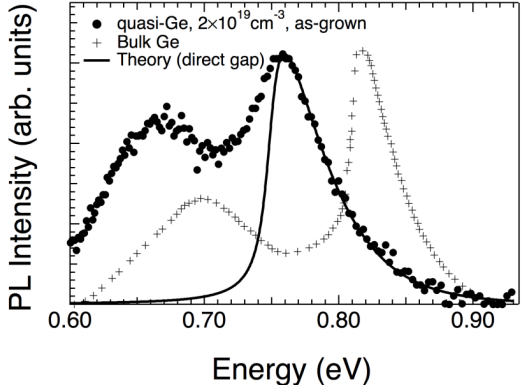

**Figure 3:** Room temperature PL spectrum of a fully relaxed, n-type Ge-on-Si film with thickness of 1200 nm and carrier density of ~ $2 \times 10^{19}/cm^3$, compared with bulk Ge emission from Haynes (Ref. 17). The spectra have been normalized to the same intensity.

The above discrepancies could be explained if the $n_\Gamma/n_L$ ratio in photoexcited *intrinsic* Ge-on-Si films deviated from the quasi-equilibrium prediction. To investigate this possibility we have carried out PL experiments in doped n-type Ge films. The rationale behind these experiments is that for doping levels close to $10^{19}$ cm$^{-3}$ the $n_\Gamma/n_L$ population ratio should be determined by standard equilibrium conditions between the two valleys and not by the photoexcitation process. Therefore, if photoexcited intrinsic films somehow deviate from quasi-equilibrium, their doped counterparts should exhibit quite different $I_{dir}/I_{ind}$ ratios. The experimental result is shown in Fig. 3 for a 1200 nm-thick Ge-on-Si film with a doping concentration of $2 \times 10^{19}$ cm$^{-1}$ and vanishing ε = 0.05%. This low level of strain is possible because the RTA step can be omitted due to the PL enhancement associated with doping. Quite remarkably, the PL spectrum lineshape now looks much more similar to the emission from bulk Ge, which is also reproduced for convenience in Fig. 3. The $I_{dir}/I_{ind}$ for the doped Ge-on-Si sample is between 1.4 and 2.1 (depending on how the background is treated). This is consistent with an enhancement of the $n_\Gamma/n_L$ ratio of about 2, as expected for this doping level (see Fig. 2), particularly if we consider the fact that the presence of donor impurities may preferentially enhance the indirect gap recombination by providing an additional channel for crystal momentum relaxation. We note that in spite of the vanishing strain, there is still a 40 meV redshift in the doped film PL relative to bulk Ge. This shift is very close to the redshift of 37 meV in the direct gap absorption edge observed by Haas for a P-doped Ge sample at a level of $1.95 \times 10^{19}$ cm$^{-3}$.[20] Haas noted that this band gap renormalization induced by the presence of P impurities is the same for the indirect and direct edges, so that it is not relevant for our discussion of $n_\Gamma/n_L$ ratios. A detailed analysis is postponed for a future publication.

The different $I_{dir}/I_{ind}$ ratios found in Ge-on-Si and bulk Ge materials, combined with the observed changes in this ratio upon doping, provide strong evidence that quasi-equilibrium conditions cannot be assumed to prevail in the conduction band of photexcited intrinsic Ge-on-Si films. We present here a very simple toy model that suggests a mechanism for the departure from equilibrium. We consider a system of two interacting states, representing the $\Gamma$-and $L$ valleys. We assume that electrons are being pumped into the $\Gamma$-valley at a rate of $G$ electrons/s. This is the case in Ge under 980 nm laser excitation. The excited electrons will redistribute themselves between the two valleys, and they will also recombine with holes via radiative and non-radiative channels. The simplest rate equations that can be written to represent this situation are:

$$\begin{aligned}\frac{dn_\Gamma}{dt} &= G - \left(\frac{1}{\tau_{\Gamma L}} + \frac{1}{\tau_\Gamma}\right)n_\Gamma + \frac{1}{\tau_{L\Gamma}}n_L \\ \frac{dn_L}{dt} &= -\left(\frac{1}{\tau_{L\Gamma}} + \frac{1}{\tau_L}\right)n_L + \frac{1}{\tau_{\Gamma L}}n_\Gamma\end{aligned} \quad , \qquad (1)$$

The transfer rates between the valleys are characterized by the time constants $\tau_{\Gamma L}$ and $\tau_{L\Gamma}$, and the constants $\tau_\Gamma$ and $\tau_L$ represent the recombination lifetimes for each valley. There is of course an additional coupled equation involving the valence band holes, but at this level of simplicity this equation is not needed for the argument we are about to make. Eq. (1) is a generalization of a two-state model introduced by Stanton and Bailey to discuss electron dynamics in GaAs and InP.[21] If the generation and recombination terms are deleted, it is easy to show[21] that for arbitrary initial occupations the system of equations has asymptotic solutions for $t \to \infty$ that imply $n_\Gamma(\infty)/n_L(\infty) = \tau_{\Gamma L}/\tau_{L\Gamma}$. Since this limit corresponds to quasi-equilibrium, the $\tau_{\Gamma L}/\tau_{L\Gamma}$ ratio must equal the ratio $n_\Gamma/n_L$ computed in Fig. 2, which leads to $\tau_{\Gamma L}/\tau_{L\Gamma} \leq 10^{-4}$ for the total electron concentrations of relevance for this discussion. If we now consider the full Eq. (1), we note that when steady-state conditions are reached (for example, under illumination with a cw laser) the time derivatives become zero and we obtain

$$\frac{n_\Gamma}{n_L} = \frac{\tau_{\Gamma L}}{\tau_{L\Gamma}} + \frac{\tau_{\Gamma L}}{\tau_L}. \qquad (2)$$

Therefore, one can expect deviations from equilibrium if the second term on the right-hand side of Eq. (2) is comparable to or greater than $10^{-4}$. Under such circumstances the relative valley populations cannot be computed using a common quasi-Fermi level. The transfer rate $\tau_{\Gamma L}$ in Ge has been measured by time-resolved inelastic light scattering, and its value is $\tau_{\Gamma L} = 1.2$ ps.[22] On the other hand, the recombination time $\tau_L$ can be associated with the minority carrier lifetime, which in bulk Ge is about 30 μs.[16] Thus for bulk Ge $\tau_{\Gamma L}/\tau_L$ is $4 \times 10^{-8}$, and the second term in Eq. (2) is much smaller than the first term. In other words, deviations from quasi-thermal equilibrium in bulk Ge are negligible, and the PL spectrum can be calculated by assuming a unique quasi-Fermi level for the conduction band.

The situation is very different in Ge-on-Si materials due to the presence of a surface and an interface. The recombination velocity $s$ at a bare Ge surface is about 140 cm/s.[23] On the other hand, the interface recombination velocity at a dislocated Si-Ge interface can be as high as $s = 4000$ cm/s.[24] Using this value for a film thickness $W = 1$ μm, the effective lifetime becomes $\tau_L = W/2s = 12$ ns, so that $\tau_{\Gamma L}/\tau_L = 10^{-4}$ and the second term in Eq. (2) is no longer negligible. For this particular value of $\tau_{\Gamma L}/\tau_L$, the ratio $n_\Gamma/n_L$ would be doubled relative to the quasi-equilibrium prediction, and the $I_{dir}/I_{ind}$ PL ratio would be enhanced proportionally. (Of course, both $n_\Gamma$ and $n_L$ decrease in absolute terms for shorter recombination lifetimes, so that the overall PL intensity is reduced, as also observed experimentally). Interestingly, from studies of $Ge_{1-y}Sn_y$ devices with a thickness $W = 300$ nm we find recombination lifetimes approaching 5 ns,[25] very close to our estimates based on interface recombination velocities. An increase in $I_{dir}/I_{ind}$ ratios as the laser photon energy was decreased was also noted by Wagner and Viña in PL experiments on highly-doped p-type Ge.[26] In their case the $I_{dir}/I_{ind}$ increase takes place precisely as the electrons become preferentially pumped into the Γ valley, the situation mimicked by Eq. (1). Thus the simple model in Eq. (2) appears to be consistent with experimental observations. Under this scenario, the relative suppression of indirect gap emission in Ge-on-Si films is at least partially due to enhanced carrier recombination rates, and the measured of $I_{dir}/I_{ind}$ PL ratios in these samples represents a useful figure of merit to monitor the success of different strategies to increase the carrier lifetimes.

In summary, the new experimental data presented here and the associated analysis indicate that the dominance of direct gap emission in Ge-on-Si films is due to the lack of significant self absorption effects and to a departure from quasi-equilibrium conditions in the conduction band, which can be traced back to a much shorter recombination lifetime than in bulk Ge. The $I_{\text{dir}}/I_{\text{ind}}$ ratio in intrinsic materials then becomes a useful figure of merit to assess the quality of the material for possible laser applications.

This work was supported by the U.S. Air Force under contract DOD AFOSR FA9550-06-01-0442 (MURI program) and by the National Science Foundation under grant DMR-0907600.